\documentclass[preprint,showpacs,preprintnumbers,amsmath,amssymb]{revtex4}

\usepackage{graphicx}
\usepackage{mathrsfs}
\usepackage{bm}
\usepackage{color,graphicx}
\usepackage{graphicx,colordvi}

\newcommand{\bel}[1]{\begin{equation}\label{#1}}

\def\be{\begin{equation}}
\def\ee{\end{equation}}
\def\bea{\begin{eqnarray}}
\def\eea{\end{eqnarray}}

\def\ms{\medskip}
\def\siml{\;\hbox{\kern.1em \lower.7ex \hbox{$\sim$} \kern-1.12em
 \raise.5ex \hbox{$<$} \kern.1em}}
\def\simg{\;\hbox{\kern.1em \lower.7ex \hbox{$\sim$} \kern-1.12em
 \raise.5ex \hbox{$>$} \kern.1em}}

\def\d{\hbox{d}}

\def\d{\hbox{d}}

\def\siml{\hbox{\kern.1em \lower.6ex \hbox{$\sim$} \kern-1.12em
          \raise.6ex \hbox{$<$} \kern.1em }}
\def\simg{\hbox{\kern.1em \lower.6ex \hbox{$\sim$} \kern-1.12em
          \raise.6ex \hbox{$>$} \kern.1em }}

\begin{document}

\title{
ATTRACTIVE INTER-PARTICLE FORCE IN VAN DER WAALS MODEL OF MULTICOMPONENT HADRON GAS IN THE GRAND CANONICAL ENSEMBLE}

\author{ Ya.\ D.\ Krivenko-Emetov}
\email{Email: krivemet@ukr.net}
\affiliation{Institute for Nuclear Research NASU, 03680 Kiev, Ukraine}

\affiliation{National Technical University of Ukraine, 03056, Kiev}

\bigskip

\date{September, 15th 2019}
\ms

\begin{abstract}

We generalize derivation of partition functions of the grand canonical ensemble for the multicomponent van der Waals gas of interacting particles by hardcore potentials to the case of the attractive large-distance mean field. The formulas obtained by the saddle point method for the thermodynamic potentials with the transparent non-relativistic limit to the case of conserving large number of particles for different gas components like neutrons and protons of nuclear matter can be used for analysis of experimental data for the particle number ratios in nucleus-nucleus collisions at high excitation energies.

\end{abstract}


\keywords{ saddle point method, multicomponent hadron gas, grand canonical ensembles, van der Waals equation of state}

\maketitle

\section{INTRODUCTION}
For the last decade, statistical models of hadron gas are actively used to describe Large Hadron Collider (LHC) data on yield of particles in nucleus-nuclear a lot of (A + A) collisions at significant energies. The van der Waals model (VdW) with allowance for hadron repulsions at short distances proved to be particularly effective among these models. This is due to the fact that taking into account the repulsion effect leads to the prevention of undesirably large values of particle number densities at high temperatures. And also to the fact that in the collisions of heavy high-energy ions in LHC a large number of different species of particles are formed. The number of these particles is not fixed. Therefore, the model uses the formalism of the Grand Canonical Ensemble(GCE) in which thermodynamic quantities depend not on the number of particles but on chemical potentials. Over the years, various versions of VdW models have been proposed and applied to fit experimental data on the ratio of  number of 
particles in A + A collisions at LHC energies. Tens or even hundreds of hadrons of different species can be a yield born depending on  collision energy. These different hadrons will be denoted by the index "i". 
Among these models we can distinguish the model proposed in \cite{gorenstein}. In this model the intrinsic volume of the i-th kind of hadrons is expressed in terms of the radius of the hard core $R_{ii}$, we the repulsion of two different species of particles "i" and "j" is expressed respectively in terms of the radius $R_{ij}$. The introduction of phenomenological parameters  $R_{ii}$ and  $R_{ij}$ markedly changes the yield of the number of particles $N_i$ and is confirmed basically by experimental results. However, the VdW model has not been properly develops in the case when there are attraction forces between  particles. Taking this fact into account would help us to describe more subtle effects in dependence of hadronic gas pressure on the density.

We can generalize this approach taking into account the attractive long-distance interaction $U_{attr}=-a n(r)$ (n-density) with smooth coordinate dependences.
 Next we do this for a multicomponent gas of particles, and so on. For multicomponent systems we have double sums that can be transformed into a multidimensional integral, having the saddle points 
$ N_1^{*},N_2^{*}$ etc. In particular, it would be good to make the transition from theory to experiment at CERN for the nucleus-nucleus collisions. First, we must do this in the case of 
a non-relativistic limit with a conserved number of particles without creating new ones (conservation of the number of particle $ N_1^{*},N_2^{*}, ...$  ...). Then we must make the transition to the relativistic nucleus-nucleus reactions 
at high energy densities. Other development applications may be related to dynamics for calculating kinetic coefficients such as viscosity, thermal conductivity, and diffusion.

\section{ONE-COMPONENT VDW GAS}
\label{sec-vdw}
We start with the standard representation for the VdW approximation of the canonical partition function  (CPF) $Z(V,T,N)$ in the canonical ensemble (CE)  \cite{landau}:
\be\label{eq1}
Z(V,T,N)  = \frac{1}{N!} \phi(T,m)^N(V-B(T)N)^N,
\ee
where $N$ is a particle number,  $m$ is a particle mass, $V$ and $T$ are volume and temperature of gas, respectively; we use the following notations \cite{gorenstein}:
\be\label{eq:14444}
\phi (m,T) =  \frac{1}{2 \pi^2} \int_0^\infty p^2 \exp{\left(-\frac{\sqrt{m^2+p^2}}{T}\right)} dp= \frac{m^2 T}{2\pi^2}K_2(m/T),
\ee
 where $K_2(z)$ is the modified Bessel function 
and second virial coefficient is given by
\be\label{eq:1555}
B(T)= \frac{1}{2}\int_0^\infty (1-\exp(-U/T)) dV, 
\ee
is a pair potential between particles, $U = \sum_{i<j} U_{ij}$. 

The pressure in the ensemble is:
\be\label{eq2}
\mathcal{P}(V,T,N)= T\frac{\partial}{\partial V} \log[Z(V,T,N)]  =  \frac{TN}{V-B(T)N} .
\ee
For the GCE partition function
$\mathcal{Z}(V,T,\mu)$ one writes \cite{feynman} (p.138)
\be\label{eq3}
\mathcal{Z}(V,T,\mu)= \sum_{N} \exp\left(\frac{\mu N}{T}\right)
Z(V,T,N)\;,
\ee
where $\mu$ is a chemical potential.
At high temperatures $ \bigtriangleup N/T = dN'$ and for $ N > N_A$ (where $N_A$ is the Avogadro constant) one can convert summation to integration:

\bea\label{eq4}
\mathcal{Z}(V,T,\mu)&=&T\int_{0}^{ \infty} \d N' 
\exp\left(\mu N'\right) Z(V,N'(T))\nonumber\\
&=&T
\int_{0}^{ \infty} \d N' 
\exp\left(\mu N'+ \log[Z(V,N'(T))]\right) = T\int_{0}^{ \infty} \d N' 
\exp\left( \Phi(N')\right).
\eea

Integration will be done 
by the saddle point method.  The saddle-point ($N^{\ast}$) is defined from the following condition: 
$$
\mu(N^*)=\mu^* = - \left(T\frac{\partial}{\partial N} \log [Z(V,T,N)]\right)_{N=N^{\ast}} -N^*(\partial \mu /\partial N )_{N=N^*} \approx
$$
\be\label{eq5}
 \approx T \left( \log (N^*/V)- \log (\phi(T,m) + \frac{ 2B(T) N^*}{V}\right)-N^* (\partial \mu /\partial N )_{N=N^*},
\ee
where $\mu^*$ is a chemical potential at the  saddle-point and
\be\label{eq7778}
 (  \partial \mu /\partial N )_{N=N^*}= T( 1/N^*+\frac{ B(T) }{V}) \approx   T(1/n^* + B(T))/V), \qquad n^*=N^*/V.
\ee
In the thermodynamic limit ($ V \rightarrow→\infty$) we have
\be\label{eq7779}
 (  \partial \mu /\partial N )_{N=N^*} \rightarrow→   0
\ee
and 
%
\be\label{eq7}
\mu^*   \rightarrow→ \mu = \left(\frac{\partial F(V,T,N)}{\partial N}\right)=T \left( \log (N^*/V)- \log (\phi(T,m) + \frac{ 2B(T) N^*}{V}\right),  
\ee
where $F(V,T,N)=-T \log [Z(V,T,N)] $ is  free energy.

In Ref.~\cite{gorenstein} only one "maximal" term was taken into accout in sum (\ref{eq3}) at the thermodynamical limit. In the present approach due to integration by single-point method we effectively take into account other terms around $N^\ast$:
%
 as usually, a small region near $N^{\ast}$ 
yields the largest contribution to the
integral over $N'$ in Eq.\ (\ref{eq4}), i.e. approximately to
the  GCE partition function (\ref{eq3}).
 %
Then, we can provide analytical integrating  in (\ref{eq4})  over $N'$
 with the saddle point $N^{\ast}$ determined
by  condition (\ref{eq5}) for any value of the chemical 
potential $\mu^*$. For this purpose, one can just solve equation (\ref{eq5}) 
after the calculation
of the free energy $F(V,T,N)$, for instance for hard-core particles of one sort
by using the method described in \cite{gorenstein}. The argument of
exponent in  Eq.\ (\ref{eq4}) can be expanded in 
$N-N^{\ast}$ up to second order that leads to  analytical result of this 
integral in terms
of the second derivative of the free energy 
$(\partial^2 F/\partial^2 N' )_{N'=N^{\ast}}$.

The pressure can be expressed via the grand canonical partition function (GCPF) by the formula \cite{kubo} (p.39)

\be\label{eq:899}
P(T,\mu, V)=T \frac{\log \mathcal{Z}(V,T,\mu) }{V}
\ee
Since in the thermodynamic limit ($ V \rightarrow→\infty$) 
 , 
\be\label{eq:8}
P(T,\mu)=T \lim_{V \to \infty} \frac{\log \mathcal{Z}(V,T,\mu) }{V}
\ee
the pressure $ P(T, \mu) $ is determined by the largest term
of the CPF
with $N = N^*$ .
Using the VdW approximation (\ref{eq1}) for CPF one finds
\be\label{eq9999} 
P(T,\mu^*) =  \lim_{V \to \infty} \frac{T}{V} \log {\left[\sqrt { \frac {  \pi  }{2}}  \frac{  A^{N^*}}{N^{*}!} \frac {   (V-B(T) N^*)^{N^*}}{ \sqrt {  ( \partial^2 \Phi /\partial N^2 )_{N=N^*}}} \right]   } ,
\ee
where $A=\phi (m,T) \exp{(\mu^* /T)}$ , $N∗ = N^*(V, T, \mu^*) $ corresponds to the saddle point, and
$ ( \partial^2 \Phi /\partial N^2 )_{N=N^*} =(N/T) ( \partial^2 \mu /\partial N^2 )_{N=N^*}+(2/T)  ( \partial \mu /\partial N )_{N=N^*}-[1/N+2B/V]=T 2B(T)/V$.

Let us show that (\ref{eq9999}) taking into account  (\ref{eq5}) leads to the result
$$
P(T,\mu^*)  \approx T\xi[1+ B(T)\xi -(\partial \mu^* /\partial N )_{N=N^*}/T)-(  \partial^2 \Phi^* /\partial N^2 )_{N=N^* }/2V\xi]  \approx
$$
\be\label{eq888}
  \approx T\xi[1+ B(T)\xi -\log {(B(T)/ V)}/2V \xi]\rightarrow→T\xi[1+ B(T)\xi] ,
\ee
where $\xi$ is defined by the transcendental equation
\be\label{eq99}
\xi   \approx  A \exp{(-2B(T)\xi+\frac{\xi }{T}  (\partial \mu^* /\partial \xi )) } \rightarrow→  A \exp{(-2B(T)\xi)}.
\ee
At  the thermodynamic limit Eq.  (\ref{eq99}) is redused to $\xi$ obtaind in Ref.~\cite{gorenstein}, 
but out of this limit we get addition terms( Eq.  (\ref{eq888}) at  the thermodynamic limit not redused to $P(T,\mu^*)$ obtaind in Ref.~\cite{gorenstein}).

Using the asymptotic bechavior of logarithm of gamma-function 

$$
\log (\Gamma (N+1))  \approx N(\log(N)-1)
$$
 it is easy to check that the values of  $N^*$  satisfying
the maximum condition of the logarithm argument in  Eq.\ (\ref{eq9999}) are given by the formula
\be\label{eq2888} 
N^* \approx  V \xi
\ee

Substitution Eq.  (\ref{eq2888})  into Eq.   (\ref{eq9999}) and using  Eq.  (\ref{eq7}) ( or   Eq. (\ref{eq99})) yields the formula  (\ref{eq888}).
It can be also presented as  Eq.\ (\ref{eq2}) with $ N = N^* (V, T, \mu) $. This demonstrates explicitly the equivalence
between canonical and grand canonical formulations at $V \to \infty $.

For the point like particles, $R = 0$ and $b = 0$,  Eq.\ (\ref{eq888})  is reduced to the ideal gas result

\be\label{eq8887777}
P(T,\mu)= P^{id}\left( T, \mu \right) = n^{id} ( T, \mu)
\ee
One can readily obtaine that definition of the
particle number density ( whith  taks into account Eq.\ (\ref{eq99}) )  are given as
\be\label{eq1000}
n= \partial P (T, \mu)/\partial \mu = \xi
\ee
We can also calculate fluctuations ( see Ref.~\cite{fed}):

\be\label{eq100011}
<(\bigtriangleup n)^2> = T/(  \partial \mu /\partial N )_{N=N^*} V^2 \sim (n^*/V)
\ee
 The formulae are obtained by the saddle point method for the thermodynamic potentials with the transparent non-relativistic limit to the case of conserving large number of particles for different gas components like neutrons and protons of nuclear matter can be used for analysis of experimental data for the particle number ratios in nucleus-nucleus collisions at high excitation energies.
 The calculations have been carried out in canonical and large canonical ensemble for a system of particles of several components by the saddle-point method. The particles interact with the hard core potentials and with relatively short-range attraction potentials (attraction radii). 
The equation of state of pressure and the set of equations for the density of particles depending on temperature and chemical potentials have been obtained. The resulting formulas coincide with the known by the form.  The use of the saddle-point method allowed, in a natural way, to obtain chemical potentials and density fluctuations $ <(\bigtriangleup n)^2>$(Eq.\ (\ref{eq7778}), (Eq.\ (\ref{eq100011})).

\section{TWO-COMPONENT VDW GAS}
\label{sec-vdww}
The procedure of taking into account the excluded volume and attraction in the VdW model in the case of a two-component hadron gas has been generalized. In the case of two species of particles "i" and "j" ($N_1$ and $N_2$ is the number of the particle species), the CPF  has the following form:
$$
Z (V, T, N_1, N_2) =\frac{1}{N_{1}!N_{2}!} \int  \prod_{l=1}^{N_1} \frac{d^3p_l^{(1)}d^3r_l^{(1)}}{(2\pi)^3} \exp{\left(-\frac{\sqrt{(m^{(1)})^2+(p_l^{(1)})^2}}{T}\right)}
$$

\be\label{eq11}
\times \int  \prod_{k=1}^{N_2} \frac{d^3p_k^{(2)}d^3r_k^{(2)}}{(2\pi)^3} \exp{\left(-\frac{\sqrt{(m^{(2)})^2+(p_k^{(2)})^2}}{T}\right)}\times\exp\left(-U_{12}/T\right)
\ee
where $m_1, N_1 (m_2, N_2)$ are, respectively,  the mass and number of particles of the 1-st (2-nd) species,


\be\label{eq12}
U ^{(12)}=  \sum \limits_{1\leq m<l\leq N_1}^{N_1} u_{11} \left(  \vert \vec{r}^{(1)}_m-\vec{r}_l^{(1)}  \vert\right)+ \sum \limits_{1\leq k<s\leq N_2}^{N_2} u_{22} \left( \vert \vec{r}^{(2)}_k-\vec{r}_s^{(2)} \vert\right) +
\sum \limits_{ m=1}^{N_1} \sum \limits_{ k=1}^{N_2}   u_{12} \left(  \vert \vec{r}^{(1)}_m-\vec{r}_k^{(2)}  \vert\right)
\ee

After integrating over the particle momenta,   Eq.\ (\ref{eq11}) is reduced to

\be\label{eq13}
Z (V, T, N_1, N_2) =\frac{1}{N_{1}!N_{2}!} [\phi (m^{(1)},T)]^{N_1} [\phi (m^{(2)},T)]^{N_2} \int  \prod_{l=1}^{N_1} d^3r_l^{(1)}
\times \int  \prod_{k=1}^{N_2} d^3r_k^{(2)}\times\exp\left(-U_{12}/T\right)
\ee
Here we use the notation

\be\label{eq14}
\phi (m ,T) =  \frac{1}{2\pi^2} \int_0^\infty p^2 \exp{\left(-\frac{\sqrt{(m)^2+(p)^2}}{T}\right)} dp= \frac{m^2 T}{2\pi^2}K_2(m ,T)
\ee
We assume that gas is not only rarefied, but its amount is rather small. And so much so that no more than one pair of particles could collide simultaneously.
 Due to the fact that free energy is an additive function, it must have the form $F=N f(T, V/N)$. Therefore the relations obtained for a small amount of gas are
 automatically valid for large quantities. Using the Mayer functions

\be\label{eq15}
-\int f^{(u,v)}(r) d^{3}r =2B(T)= 4\pi \int_0^\infty (1-\exp(-U^{(u,v)}/T))r^2 dr  
\ee
and real form of the potential ($u,v=1,2$)
\[
\delimiterfactor=1200 
U^{(uv)} = \left\{%
\begin{array}{ll}
\infty &\textrm{if } r< R^0_{u}+R^0_{v},\\
-u_{0}^{(u,v)} &\textrm{if } R^0_{u}+R^0_{v}\le r<R_{u}+R_{v},\\
0 &\textrm{if }R_{u}+R_{v}\le r.
\end{array}%
\right.
\]
${}$
one can rewrite the $exp(-U/T)$ in Eq.\ (\ref{eq13})  in the following form

$$
\exp(-U/T)= \prod_{k=1}^{N_w-1} \prod_{l=k+1}^{N_w}[1+f_{kl}^{ww}]\prod_{i=1}^{N_v} (\prod_{j=i+1}^{N_v}[1+f_{ij}^{vv}]\prod_{m=1}^{N_w} [1+f_{im}^{vw}])
$$

\be\label{eq16}
\approx\prod_{k=1}^{N_w-1} [1+\sum_{l=k+1}^{N_w}f_{kl}^{ww}]\prod_{i=1}^{N_v} (1+\sum_{j=i+1}^{N_v}f_{ij}^{vv}+\sum_{m=1}^{N_w}f_{im}^{vw})
\ee
where we use
$$
\int f_{ij}f_{kl} dr_i dr_j dr_k dr_l 
\approx \int (u_{ij}u_{kl}/T^2) dr_i dr_j dr_k dr_l->0
$$
and
$
f_{ij}f_{kl}f_{im}->0
$.
The assumption that the gas is sufficiently rarefied allows us to neglect the triple interaction $ U_{123}<< U_{12}$  and to use the high temperature condition $U_{12}/T<< 1$. Then

$$
4\pi [\int_0^{R^0_{u}+R^0_{v}} r^2 dr +\int_{R^0_{u}+R^0_{v}}^{R_{v}+R_{u}} (1-\exp(u_0^{(uv)}/T))r^2 dr]=
$$

\be\label{eq:17}
= \frac{4}{3}\pi (R^0_{u}+R^0_{v}) ^3 +\frac{4}{3}\pi [-(R^0_{u}+R^0_{v})^3+(R_{u}+R_{v})^3] (1-\exp(u_0^{(uv)}/T)) 
\ee

\be\label{eq:18}
\approx  \frac{4}{3}\pi (R^0_{u}+R^0_{v}) ^3- (u_0^{(uv)}/T) \frac{4}{3}\pi [(R_{u}+R_{v})^3-(R^0_{u}+R^0_{v})^3]=2(b_{uv}-a_{uv}/T)=2B_{uv}(T)
\ee

where constants are denoted as $b_{uv}=\frac{2}{3}\pi (R^0_{u}+R^0_{v})^3,   \qquad   c_{uv}=\frac{2}{3}\pi (R_{u}+R_{v})^3$
and
$a_{uv}=u^{(uv)}_0 (c_{uv}-b_{uv})$.

Substituting Eqs.\ (\ref{eq14}) - Eqs.\ (\ref{eq16}) into Eq.\ (\ref{eq13})   and imposing the additional constraints $ 2N B/V << 1$, one obtains the final expression 
for the partition function of the two-component VdW gas:

$ Z(V,T,N_1,N_2)  \sim$
\be\label{eq:19}
 \sim \frac{1}{N_1!N_2!} \phi(T,m_1)^N_{1} \phi(T,m_2)^N_{2}(V-B_{11}N_1-\tilde{B}_{21}N_2)^{N_1}(V-B_{22}N_2-\tilde{B}_{12}N_1)^{N_2}
\ee

where

$
 \tilde{B}_{ij}=2 \frac{B_{ii}B_{ij}}{B_{ii}+B_{jj}} \approx  \tilde{b}_{ij}- \tilde{a}_{ij}/T
$,
$
 \tilde{b}_{ij}=2 \frac{b_{ii}b_{ij}}{b_{ii}+b_{jj}}
$,
and

\be\label{eq20001}
\tilde{a}_{ij}  \approx a_{ij} b_{ii}/(b_{ii}+b_{jj}) -b_{ij}( b_{ii}a_{ii}+2b_{ii}a_{jj}-b_{jj}a_{ii})/(b_{ii}+b_{jj})^2
\ee

Thus, the use of the potential  $U^{uv}$ Eq.\ (\ref{eq12})  made it possible to obtain relations Eq.\ (\ref{eq20001}) by comparison with article  \cite{gorvov}.

Substituting this expression into the formula for the pressure and using  $a_{ij}N_iN_j/V^2<<1$ one obtains

$$
\mathcal{P}(V,T,N_1,N_2)= T\frac{\partial}{\partial V} Log[Z(V,T,N_1,N_2)]  \sim \frac{TN_1}{V-B_{11}N_1-\tilde{B}_{21}N_2}+\frac{TN_2}{V-B_{22}N_2-\tilde{B}_{12}N_1}
$$

\be\label{eq:20}
\approx  \frac{TN_1}{V-b_{11}N_1-\tilde{b}_{21}N_2}+\frac{TN_2}{V-b_{22}N_2-\tilde{b}_{12}N_1} - \frac{N_1 (a_{11} N_1+\tilde{a}_{21} N_2)}{V^2}-\frac{N_2 (a_{22} N_2+\tilde{a}_{12} N_1)}{V^2}
\ee

Using condishions $b_{ij}N_iN_j/V^2<<1$  the last  formula can be rewritten in the familiar form of the Van der Waals virial decomposition



\be\label{eq:202}
\mathcal{P}(V,T,N_1,N_2)   \approx \frac{TN_1}{V}+\frac{TN_2}{V}+\frac{B_{11}TN_1^2}{V^2}+\frac{(B_{12}+B_{21})TN_1N_2}{V^2}+ \frac{B_{22}N^2_2}{V^2}
\ee

For  the GCE partition function one writes
\be\label{eq:21}
\mathcal{Z}(V,T,\mu_1,\mu_2)  =  \sum_{N_1,N_2} \exp\left(\frac{\mu_1 N_1+\mu_2 N_2}{T}\right)
Z(V,T,N_1,N_2) 
\ee
Using the saddle point method and Eq.\ (\ref{eq4}), one obtains

\be\label{eq:22}
P(T,\mu_1,\mu_2)  \sim \lim_{V \to \infty} \frac{T}{V} \log  \sqrt { \frac {  \pi}{ 2  [ Log( Z)]^{''}}}        \exp\left(\frac{\mu_1 N_1^*+\mu_2 N_2^*}{T}\right)
Z(V,T,N_1^*,N_2^*) 
\ee

Here $N^*$ is also the average number of particles in the grand canonical formulation.
Using the VdW approximation  Eq.\ (\ref{eq:19}) for the CPF one finds

$$
P(T,\mu_1,\mu_2)  \sim
$$
\be\label{eq:23}
 \sim \lim_{V \to \infty} \frac{T}{V} \log {   \left( \sqrt { \frac {  \pi}{ 2   [Log( Z)]^{''}}}  \frac{1}{N^*_1!N^*_2!} A_1^{N^*_{1}} A_2^{N^*_{2}} (V-B_{11}N^*_1-\tilde{B}_{21}N^*_2)^{N^*_1}(V-B_{22}N^*_2-\tilde{B}_{12}N^*_1)^{N^*_2}\right)}
\ee
($A_i=\phi (m_i,T) \exp{(\mu_i /T)}$)

Let us show that the pressure  can be calculated by the formula \cite{gorenstein}, \cite{krivenko}:

\be\label{eq:25}
P(T,\mu_1,\mu_2)  \sim T[\xi_1+\xi_2  +\xi_1^2 B_{11}+\xi_2^2 B_{22} + (B_{12}+B_{21})\xi_1\xi_2 ]
\ee

where the values of $ \xi_q$ are found from the set of coupled transcendental equations

\be\label{eq:26}
\xi_1  \sim A_1 \exp{(-2B_{11}\xi_1-2\tilde{B}_{12}\xi_2)}
\ee

\be\label{eq:27}
\xi_2  \sim A_2 \exp{(-2B_{22}\xi_2-2\tilde{B}_{21}\xi_1)}
\ee

Using the asymptotic representation for the gamma-function logarithm

$$
log (\Gamma (N+1)) \sim N(log(N)-1)
$$
 it is easy to check that the values of  $N^*_{1,2}$  satisfying
the maximum condition of the logarithm argument in  Eq.\ (\ref{eq:23}) are given by the formula
\be\label{eq:28} 
N^*_{1,2} \approx  V n_{1,2}
\ee

where $n_p = n_p(T, \mu_1, \mu_2)$ are related to $\xi_p$ via the equations

\be\label{eq:31}
\xi_1 \sim n_1
\ee
\be\label{eq:32}
\xi_2  \sim n_2
\ee
Substituting Eq.\ (\ref{eq:31}) and Eq.\ (\ref{eq:32}) into Eq.\ (\ref{eq:25}) ,we obtain the VdW equation of state in the GCE:

\be\label{eq:33}
P(T,\mu_1,\mu_2)   \sim \frac{Tn_1}{1-b_{11}n_1-\tilde{b}_{21}n_2}+\frac{Tn_2}{1-b_{22}n_2-\tilde{b}_{12}n_1} -  n_1 (a_{11} n_1+\tilde{a}_{21} n_2)-n_2 (a_{22} n_2+\tilde{a}_{12} n_1)
\ee

\be\label{eq:34}
P(T,\mu_1,\mu_2)   \sim \frac{Tn}{1-b_{11}n-\tilde{b}_{21}n}+\frac{Tn}{1-b_{22}n-\tilde{b}_{12}n} -  (a_{11}+\tilde{a}_{21} + \tilde{a}_{12}+a_{22}) n^2 \sim  \frac{T2n}{1-b 2n} -a (2n)^2 
\ee
where
$b=[ b_{11}+b_{22}+2b_{12}]/2$, and $a=(a_{11}+\tilde{a}_{21} + \tilde{a}_{12}+a_{22})/4 $

Calculate the critical points of gas VdW:

\be\label{eq:35}
T_c = 8 a n_c/[\frac{1}{(1-b_1n_c)^2}+\frac{1}{(1-b_2n_c)^2}] =  \frac{8a}{27b_{11}}
\ee
at the simmetric $(1-b_1n_c) = 2/3$ and $T_c= \frac{8a}{27b_{11}}$
\be\label{eq:36}
[\frac{1}{(1-b_1n_c)^3}+\frac{1}{(1-b_2n_c)^3}] =\frac{3}{2}[\frac{1}{(1-b_1n_c)^2}+\frac{1}{(1-b_2n_c)^2}] 
\ee

\be\label{eq:37}
n_c \sim \sqrt { \frac{2}{9(b_1^2+b_2^2)}}
\ee

if $b_{11}=b_{22}$, $ b_1 =b_2=2b_{11}=2b_{22}$

\be\label{eq:38}
n_c =N_c/2 \sim \sqrt { \frac{1}{9*4(b_{11}^2)}} 
\ee
%
and
\be\label{eq:41}
P_c =   \frac{a}{27b_{11}^2} 
\ee

\section{CONCLUSION}

In the paper the effect of taking into account the excluded volume and attraction in the VdW model in the case of a two-component hadronic gas is analyzed. The calculations have been carried out in canonical and large canonical ensemble for a system of particles of several components by the saddle-point method. The particles interact with the hard core potentials and with relatively short-range attraction potentials (attraction radii). 
The equation of state of pressure and the set of equations for the density of particles depending on temperature and chemical potentials have been obtained. The resulting formulas coincide with the known by the form. From the latter, they differ in the content of new components  $\tilde{a}_{ij} $(Eq.\ (\ref{eq20001}) ). The use of the saddle-point method allowed, in a natural way, to obtain pressure, chemical potentials and density fluctuations $ <(\bigtriangleup n)^2>$(Eq.\ (\ref{eq888}), Eq.\ (\ref{eq:25}), Eq.\ (\ref{eq7778}), (Eq.\ (\ref{eq100011})). 

The obtained formulas are consistent with the basic principles of statistical mechanics, as well as with thermodynamic identities. The developed model can be applied for the analysis of experimental data for the relative yield of particles of different species in relativistic nuclear-nuclear collisions.

\section{Acknowledgements}
I sinserely thank Prof. A.P. Kobushkin for fruitful discussion and helpful comments.

\section{Figures}
%
%
 \begin{center}
\includegraphics{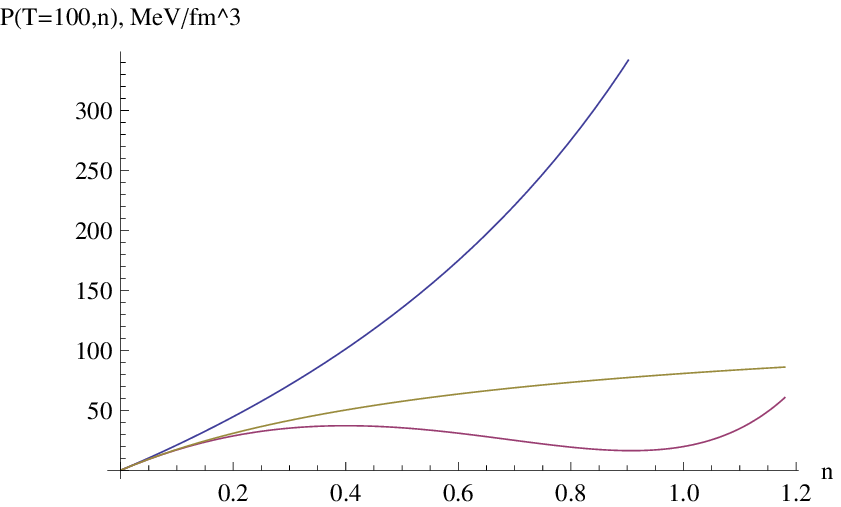}
 \end{center}
 Fig. 1. Pressure for real radii nuclons, dependence on density calculated for an ideal gas and for models corresponding to the model of \cite{gorenstein}  and Eq.\ (\ref{eq:33}) (at T=100 MeV)
\begin{center}
 \includegraphics{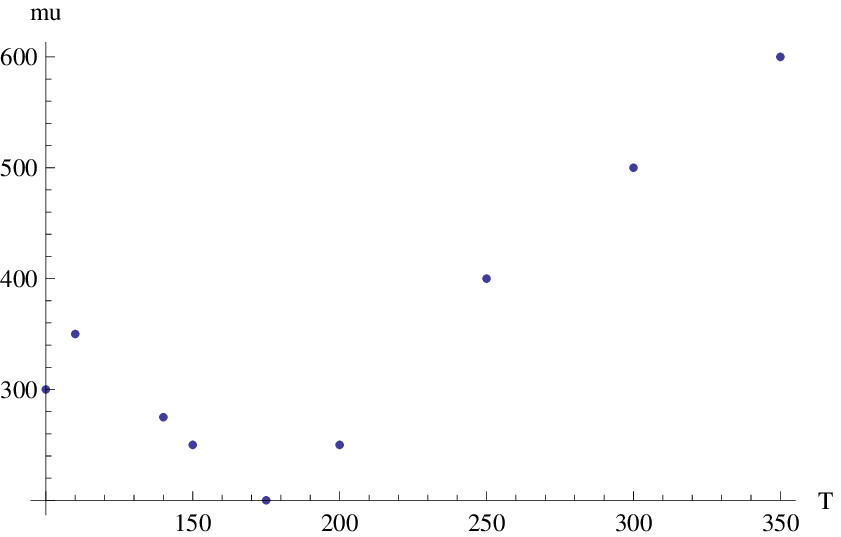}
 \end{center}
Fig 2. Chemical potential dependence of temperature [MeV]
%
%
%

\end{document}